%% file: Identifying_Non-autonomous_Dynamical_System_with_Physics-Guided_Recurrent_Neural_Networks.tex
\documentclass{ifacconf}

\usepackage{graphicx}      
\usepackage{natbib}        

\usepackage{amsmath,amssymb,amsfonts}
\usepackage{textcomp}
\usepackage{xcolor}
\usepackage{subfig}
\renewcommand{\vec}{\boldsymbol}
\DeclareMathAlphabet{\mathscr}{U}{BOONDOX-cal}{m}{n}
\SetMathAlphabet{\mathscr}{bold}{U}{BOONDOX-cal}{b}{n}
\DeclareMathAlphabet{\mathbscr} {U}{BOONDOX-cal}{b}{n}
\newcommand{\rec}[1]{\stackrel{\curvearrowleft}{#1}}
\usepackage{siunitx}  
\definecolor{hniblue}{RGB}{51, 145, 202}
\definecolor{hnired}{RGB}{219, 72, 72}
\definecolor{hnigray}{RGB}{105, 106, 107}
\definecolor{hnilightgreen}{RGB}{105, 186,179}
\usepackage{wrapfig}
\newcommand*\widebar[1]{%
	\hbox{%
		\vbox{%
			\hrule height 0.5pt 
			\kern0.5ex
			\hbox{%
				\kern-0.1em
				\ensuremath{#1}%
				\kern-0.1em
			}%
		}%
	}%
} 
\begin{document}
\begin{frontmatter}

\title{Multi-Objective Physics-Guided Recurrent Neural Networks for Identifying Non-Autonomous Dynamical Systems\thanksref{footnoteinfo}}

\thanks[footnoteinfo]{This work was developed in the junior research group DART (\emph{Daten\-ge\-trie\-be\-ne Methoden in der Regelungstechnik}), University Paderborn, and funded by the German Federal Ministry of Education and Research BMBF (\emph{Bundesministerium f\"ur Bildung und Forschung}) under the funding code 01IS20052. The responsibility for the content of this publication lies with the authors. \copyright{} 2022 the authors. This work has been accepted to IFAC for publication under a Creative Commons Licence CC-BY-NC-ND.}

\author{Oliver Sch\"on, Ricarda-Samantha G\"otte, Julia Timmermann}

\address{Heinz Nixdorf Institute,
	Paderborn University, Germany, (e-mail: \{oliver.schoen, rgoette, julia.timmermann\}@ hni.upb.de)}

\begin{abstract}                
While trade-offs between modeling effort and model accuracy remain a major concern with system identification, 
resorting to data-driven methods often leads to a complete disregard for physical plausibility.
To address this issue, we propose a physics-guided hybrid approach for modeling non-autonomous systems under control. Starting from a traditional physics-based model, this is extended by a recurrent neural network and trained using a sophisticated multi-objective strategy yielding physically plausible models. 
While purely data-driven methods fail to produce satisfying results, experiments conducted on real data reveal substantial accuracy improvements by our approach compared to a physics-based model.
\end{abstract}


\end{frontmatter}

\section{Introduction}
Modeling real world systems usually involves a certain level of abstraction. Models derived from fundamental physics laws, for example, in the remainder referred to as \emph{physics-based models}, are one of the most prevalent type of model in engineering, e.g., in form of ordinary differential equations (ODEs). These abstract representations of an underlying real system only capture a limited amount of its dynamics, omitting statistical aspects or complex nonlinear effects, either compromising model quality due to modeling effort or due to a system's complexity even with high modeling effort.

Not all aspects in the process of modeling a technical system can be theoretically derived. Determining the model parameters, for example, usually involves experimental identification, which is why measurement data is often collected. To further reduce the modeling effort and mitigate the effects of insufficient system understanding, purely data-driven approaches are increasingly being adopted in modern engineering disciplines, e.g., see \cite{Brunton.2016, Junker.2021}. However, while the resulting models might be suitable for some applications, they are completely unfit for settings where actions need to be explainable and verifiably valid, see \cite{Garcez.2020}, especially important in safety-critical domains like autonomous transport or healthcare. Additional limitations include poor extrapolative properties as well as lacking compatibility with the most common engineering tools (e.g., stability analysis). To address these deficiencies, we introduce a new class of hybrid models.

Since their introduction by \cite{Karpatne.2017}, Physics-Guided Neural Networks (PGNNs), a novel type of hybrid models, have already been successfully implemented in several domains of application, e.g., see \cite{Karpatne.2017b, Willard.2021}. As a result, by embedding physics-based models and restrictions as in \cite{Dener.2020}, both synergetic effects as well as physically sound models were obtained. Based on prior works, c.f. \cite{Antonelo.2021, Gotte.2022}, the potential of PGNNs for the identification of non-autonomous dynamical systems is revisited from a control engineering point of view using a refined approach.

By combining a recurrent neural network with a physics-based model, a Physics-Guided Recurrent Neural Network (PGRNN) is constructed. Furthermore, physics-based constraints are introduced, leading to a Multi-Objective Physics-Guided Recurrent Neural Network (MOPGRNN). It is demonstrated that the MOPGRNN outperforms a purely data-driven approach by a substantial margin.

Our contributions to the field of physics-guided models are the following:
\begin{itemize}
	\item[(1)] Increased model accuracy through a combined recurrent physics-based modeling approach for non-autonomous systems under control (i.e., PGRNN, Sec.~\ref{sec:pgrnn}),
	\item[(2)] enforcing physics-based restrictions using a sophisticated multi-objective strategy, inciting physically plausible models whilst securing training conditioning (i.e., MOPGRNN, Sec.~\ref{sec:physics_based_constraints}),	
	\item[(3)] introducing different objective functions for training vis-\`{a}-vis validation and testing, combining differentiable and non-differentiable objective functions (Sec.~\ref{sec:simulative_experiments}).
\end{itemize}

The paper is outlined as follows: In Sec.~\ref{sec:method} we extend our previous results from \cite{Gotte.2022} and propose a MOPGRNN to obtain physically plausible models. To demonstrate its advantages for system identification in terms of increased accuracy and plausibility, two nonlinear systems are studied in Sec.~\ref{sec:experiments}, including a real application system. A short discussion of the results concludes the paper (Sec.~\ref{sec:conclusion}). 

\section{Main idea}\label{sec:method}

When modeling systems, we often face the following situations. Either a simplistic physics-based model can be constructed with comparatively low effort or a complex model is developed with high effort but still lacks of sufficient accuracy. Using such models as a basis, the main idea of this work is to learn a mapping function in a data-driven manner, transforming low-quality physics-based dynamics to high-quality dynamics. Therefore, we constitute a \emph{Physics-Guided Recurrent Neural Network} as in \cite{Jia.2020}, using a set of ODEs as a physics-based model and a recurrent neural network (RNN) as the data-driven mapping component. Note that \cite{Jia.2020} only consider autonomous systems with a single objective function. By utilizing a simplistic physics-based model we are hoping for synergetic effects, e.g., lower data requirements and increased model accuracy. Furthermore, by enforcing physics-based constraints  on the mapping function in a multi-objective optimization setting, physically plausible hybrid models are expected to be retrieved.

In the course of modeling time-variant, non-autonomous systems, we first take a look at the specification of feed-forward and recurrent neural networks. Then, by combining a RNN and a physics-based model, we derive the PGRNN and augment it using physics-based constraints, yielding a \emph{Multi-Objective Physics-Guided Recurrent Neural Network}.

Notationally, to differentiate variables describing a real system $\mathcal{S}$ and different model types, we denote variables associated with a physics-based model with a superscript $\tilde{\cdot}$, and (partially) data-driven models with $\hat{\cdot}$ or swung letters. 

\subsection{Feed-Forward Neural Network}\label{sec:nn}
The most prevalent form of Neural Networks (NNs) are feed-forward NNs, therefore usually simply referred to as neural networks. They consist of several layers of neurons and forward-facing connections between them, functionally transforming an input $\vec{x} \in \mathbb{X} \subseteq \mathbb{R}^n$ into an output signal $\hat{\vec{y}} \in \mathbb{Y} \subseteq \mathbb{R}^{l}$, often used to recognize and model patterns in data. A NN $\vec{\mathscr{f}}$ is defined by 
\begin{equation}
\hat{\vec{y}} = \vec{\mathscr{f}}\left( \vec{x} ; \vec{\theta} \right), \quad \vec{\mathscr{f}}: \mathbb{X} \mapsto \mathbb{Y},
\end{equation}
and its learnables $\vec{\theta}:=\left[  (\vec{w}_i,\vec{b}_i); i=1,\ldots \right] \in \mathbb{R}^{\mathscr{q}}$ with weight parameters $\vec{w}_i$ of the connections and biases $\vec{b}_i$ of the neurons for each layer $i$.
Through an iterative training scheme, the learnables are adjusted to fit the network's output $\hat{\vec{y}}$ to a given target $\vec{y} \in \mathbb{Y}$ by error backpropagation, c.f. \cite{Bishop.2006}. Besides the mentioned learnable parameters, a NN is determined by a comprehensive set of hyperparameters. These include, e.g., the number of hidden neurons, characterizing the complexity of the networks modeling capability.	

Let the dynamics and output functions $\vec{f}$ and $\vec{g}$ of a time-variant, non-autonomous system $\mathcal{S}$ with state $\vec{x}_{t} \in \mathbb{X} \subseteq \mathbb{R}^{n}$, control input $\vec{u}_{t} \in \mathbb{U} \subseteq \mathbb{R}^{m}$, output $\vec{y}_{t} \in \mathbb{Y} \subseteq \mathbb{R}^{l}$, evolving in time $t \in \mathbb{T} \subseteq \mathbb{R}$, as well as a set of parameters $\vec{p} \in \mathbb{R}^{q}$ be given by
\begin{align}
\begin{alignedat}{2}
\dot{\vec{x}}_{t} &= \vec{f}\left( \vec{x}_{t}, \vec{u}_{t}, t ; \vec{p} \right), \quad & \vec{f} &: \mathbb{X} \times \mathbb{U} \times \mathbb{T} \mapsto \mathbb{X},\\
\vec{y}_{t} &= \vec{g}\left( \vec{x}_{t}, \vec{u}_{t}, t ; \vec{p} \right), \quad & \vec{g} &: \mathbb{X} \times \mathbb{U} \times \mathbb{T} \mapsto \mathbb{Y}.
\end{alignedat}
\end{align}

Then, we can model the system using a NN $\mathcal{M}_{NN}$, comprising the dynamics and output functions $\vec{\mathscr{f}}$ and $\vec{\mathscr{g}}$, specified by a state $\hat{\vec{x}}_{t} \in \hat{\mathbb{X}} \subseteq \mathbb{R}^{\hat{n}}$, output $\hat{\vec{y}}_{t} \in \mathbb{Y}$ and learnables $\vec{\theta}$:
\begin{equation}
\begin{alignedat}{2}
\dot{\hat{\vec{x}}}_{t}  &= \vec{\mathscr{f}}\left( \hat{\vec{x}}_{t}, \vec{u}_{t}, t ; \vec{\theta} \right), \quad & \vec{\mathscr{f}} &: \hat{\mathbb{X}} \times \mathbb{U} \times \mathbb{T} \mapsto \hat{\mathbb{X}},\\
\hat{\vec{y}}_{t}  &= \vec{\mathscr{g}}\left( \hat{\vec{x}}_{t}, \vec{u}_{t}, t ; \vec{\theta} \right), \quad & \vec{\mathscr{g}} &: \hat{\mathbb{X}} \times \mathbb{U} \times \mathbb{T} \mapsto \mathbb{Y}.
\end{alignedat}
\end{equation}

\subsection{Recurrent Neural Network}\label{sec:rnn}
The most common type of data acquired for modeling dynamical systems is sequential data, implying temporal dependencies between subsequent data samples. RNNs are particularly equipped to model long-term temporal dependencies by introducing a hidden state $\vec{h}_k \in \mathbb{H} \subseteq \mathbb{R}^{z}$, and modeling the information flow into and out of the hidden state using so-called gates. Here, $z$ is the number of hidden neurons, which is usually much higher than the number of inputs. 

Input to a RNN is supplied in form of sequential data comprising the past and current input, here referred to as input history. With the state history $\mathcal{H}^{\hat{\vec{x}}}_{t} := \hat{\vec{x}}_{t_0},\ldots,\hat{\vec{x}}_{t}$ and control input history $\mathcal{H}^{\vec{u}}_{t} := \vec{u}_{t_0},\ldots,\vec{u}_{t}$, each containing $N_t$ samples, as well as the corresponding time history $\mathcal{H}^{t}_{t} := t_0,\ldots,t$ being equally distributed, we obtain a model $\mathcal{M}_{RNN}$, defined by \eqref{eq:RNN}, where the dynamics and output functions are modeled by two RNNs $\rec{\vec{\mathscr{f}}}$ and $\rec{\vec{\mathscr{g}}}$, respectively, the backward arrow indicating their recurrent nature:
\begin{align}
\begin{alignedat}{2}
\dot{\hat{\vec{x}}}_{t} &= 	\,\rec{\vec{\mathscr{f}}}\left( \mathcal{H}^{\hat{\vec{x}}}_{t}, \mathcal{H}^{\vec{u}}_{t}, \mathcal{H}^{t}_{t} ; \vec{\theta} \right), \, & \,\rec{\vec{\mathscr{f}}} &: \hat{\mathbb{X}}^{N_t} \times \mathbb{U}^{N_t} \times \mathbb{T}^{N_t} \mapsto \hat{\mathbb{X}},\\
\hat{\vec{y}}_{t} &= \,\rec{\vec{\mathscr{g}}}\left( \mathcal{H}^{\hat{\vec{x}}}_{t}, \mathcal{H}^{\vec{u}}_{t}, \mathcal{H}^{t}_{t} ; \vec{\theta} \right), \, & \,\rec{\vec{\mathscr{g}}} &: \hat{\mathbb{X}}^{N_t} \times \mathbb{U}^{N_t} \times \mathbb{T}^{N_t} \mapsto \mathbb{Y}.
\end{alignedat}\label{eq:RNN}
\end{align}

Each RNN consists of two components, a Gated Recurrent Unit (GRU), c.f. \cite{Cho.2014}, and a single layer of neurons. 
The GRU is used to model the information flow between subsequent hidden states $\vec{h}_k$, $k = 0, \ldots, N_t$, taking the inputs $\vec{s}_k = \left[ \hat{\vec{x}}_{t_k}, \vec{u}_{t_k}, t_k \right]^{T}$ into account. It is defined by learnable network parameters, comprising input weights $\vec{W}$, recurrent weights $\vec{R}$ and biases $\vec{b}$:
\begin{align}
\vec{W} := \begin{bmatrix} \vec{W}_{\vec{z}} \\ \vec{W}_{\vec{r}} \\ \vec{W}_{\vec{\tilde{h}}} \end{bmatrix},\quad
\vec{R} := \begin{bmatrix} \vec{R}_{\vec{z}} \\ \vec{R}_{\vec{r}} \\ \vec{R}_{\vec{\tilde{h}}} \end{bmatrix},\quad
\vec{b} := \begin{bmatrix} \vec{b}_{\vec{z}} \\ \vec{b}_{\vec{r}} \\ \vec{b}_{\vec{\tilde{h}}} \end{bmatrix}.
\end{align}

An update gate $\vec{z}_k$ is used to add new information to the hidden state $\vec{h}_k$, which acts like a temporal memory, while a reset gate $\vec{r}_k$ is used for successive removal of information, interpretable as an act of forgetting. Using these, a candidate state $\vec{\tilde{h}}_k$ is calculated:
\begin{align}
\begin{split}
\vec{z}_k &= \vec{\sigma}_g\left( \vec{W}_{\vec{z}}\vec{s}_k + \vec{b}_{\vec{z}} + \vec{R}_{\vec{z}}\vec{h}_{k-1} \right),\\
\vec{r}_k &= \vec{\sigma}_g\left( \vec{W}_{\vec{r}}\vec{s}_k + \vec{b}_{\vec{r}} + \vec{R}_{\vec{r}}\vec{h}_{k-1} \right),\\
\vec{\tilde{h}}_k &= \vec{\sigma}_s\left( \vec{W}_{\tilde{\vec{h}}}\vec{s}_k + \vec{b}_{\vec{\tilde{h}}} + \vec{r}_k \odot \left( \vec{R}_{\tilde{\vec{h}}}\vec{h}_{k-1} \right)  \right).
\end{split}\label{eq:gru}
\end{align}

The new hidden state $\vec{h}_k$ results as a linear combination of the preceding hidden state $\vec{h}_{k-1}$ and the candidate state $\vec{\tilde{h}}_k$, i.e., $\vec{h}_k = \left( 1 - \vec{z}_k \right) \odot \vec{h}_{k-1} + \vec{z}_k \odot \tilde{\vec{h}}_k$. Here, the gate activation function $\vec{\sigma}_g$ and state activation function $\vec{\sigma}_s$ are the logistic and tanh-function, respectively, while $\odot$ indicates element-wise multiplication. The GRU's output at time $t$ is the current hidden state $\vec{h}_{N_t}$. Following up the GRU with a consecutive layer of neurons, the high dimensional hidden state $\vec{h}_{N_t}$ is transformed back into state space $\hat{\mathbb{X}}$ and output space $\mathbb{Y}$, i.e., $\dot{\hat{\vec{x}}}_{t} := \vec{w}_b^{T}\vec{h}_{N_t} + \vec{b}_b$, $\vec{w}_b$ and $\vec{b}_b$ being the consecutive layer's weights and biases. The output $\hat{\vec{y}}_{t}$ is retrieved analogously.

\subsection{Physics-Guided Recurrent Neural Network}\label{sec:pgrnn}
Let a physics-based model $\mathcal{M}_{phy}$ with state $\tilde{\vec{x}} \in \tilde{\mathbb{X}} \subseteq \mathbb{R}^{\tilde{n}}$, output $\tilde{\vec{y}} \in \mathbb{Y}$ and parameters $\tilde{\vec{p}} \in \mathbb{R}^{\tilde{q}}$ be given by
\begin{align}
\begin{alignedat}{2}
\dot{\tilde{\vec{x}}}_{t} &= \tilde{\vec{f}}\left( \tilde{\vec{x}}_{t}, \vec{u}_{t}, t ; \tilde{\vec{p}} \right), \quad & \tilde{\vec{f}} &: \tilde{\mathbb{X}} \times \mathbb{U} \times \mathbb{T} \mapsto \tilde{\mathbb{X}},\\
\tilde{\vec{y}}_{t} &= \tilde{\vec{g}}\left( \tilde{\vec{x}}_{t}, \vec{u}_{t}, t ; \tilde{\vec{p}} \right),\quad & \tilde{\vec{g}} &: \tilde{\mathbb{X}} \times \mathbb{U} \times \mathbb{T} \mapsto \mathbb{Y},
\end{alignedat}
\end{align}
modeling the reference system $\mathcal{S}$ approximately. Then, the model $\mathcal{M}_{PGRNN}$, relating physics-based dynamics $\dot{\tilde{\vec{x}}}_{t} \in \mathbb{X}$ and outputs $\tilde{\vec{y}}_{t} \in \mathbb{Y}$ to $\dot{\hat{\vec{x}}}_{t} \in \hat{\mathbb{X}}$ and $\hat{\vec{y}}_{t} \in \mathbb{Y}$, respectively, is given by two RNNs $\rec{\vec{\mathscr{F}}}$ and $\rec{\vec{\mathscr{G}}}$ modeling the dynamics and output functions:
\begin{align}
\begin{split}
\dot{\hat{\vec{x}}}_{t} &= \,\rec{\vec{\mathscr{F}}}\left( \mathcal{H}^{\hat{\vec{x}}}_{t}, \mathcal{H}^{\vec{u}}_{t}, \mathcal{H}^{t}_{t}, \mathcal{H}^{\dot{\tilde{\vec{x}}}}_{t} ; \vec{\theta} \right),\\
\hat{\vec{y}} &= \,\rec{\vec{\mathscr{G}}}\left( \mathcal{H}^{\hat{\vec{x}}}_{t}, \mathcal{H}^{\vec{u}}_{t}, \mathcal{H}^{t}_{t}, \mathcal{H}^{\tilde{\vec{y}}}_{t} ; \vec{\theta} \right),\\ 
\,\rec{\vec{\mathscr{F}}}&: \hat{\mathbb{X}}^{N_t} \times \mathbb{U}^{N_t} \times \mathbb{T}^{N_t} \times \tilde{\mathbb{X}}^{N_t} \mapsto \hat{\mathbb{X}},\\
\,\rec{\vec{\mathscr{G}}}&: \hat{\mathbb{X}}^{N_t} \times \mathbb{U}^{N_t} \times \mathbb{T}^{N_t} \times \mathbb{Y}^{N_t} \mapsto \mathbb{Y},
\end{split}
\end{align}
where $\mathcal{H}^{\dot{\tilde{\vec{x}}}}_{t} := \dot{\tilde{\vec{x}}}_{t_0},\ldots,\dot{\tilde{\vec{x}}}_{t}$ and $\mathcal{H}^{\tilde{\vec{y}}}_{t} := \tilde{\vec{y}}_{t_0},\ldots,\tilde{\vec{y}}_{t}$ denote the physics-based dynamics and output histories, respectively.
The RNNs are trained to model target dynamics $\dot{\vec{x}}_{t}$ and outputs $\vec{y}_{t}$ by transforming physics-based information in a data-driven fashion to enhance the accuracy of predictions and lower training requirements. Therefore, in analogy to \cite{Jia.2020}, the model is called a \emph{Physics-Guided Recurrent Neural Network}. Note that we are considering the controlled case here, i.e., augmented with $\mathcal{H}^{\vec{u}}_{t}$.

\begin{figure}[h]
	\centering
	\def\svgwidth{\columnwidth}	
	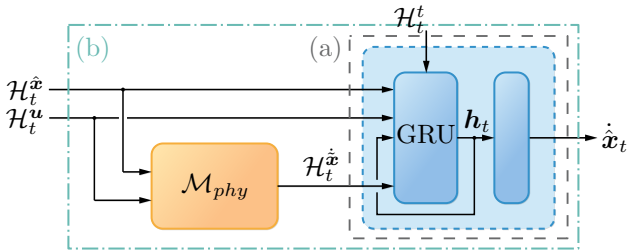
	\caption{Two interpretations of the PGRNN, either (a) $\mathcal{M}_{PGRNN}$ or (b) $\widebar{\mathcal{M}}_{PGRNN}$}
	\label{fig:pgrnn}
\end{figure}
By shifting the model's boundaries (Fig.~\ref{fig:pgrnn}), the physics-based model $\mathcal{M}_{phy}$ can be embedded within the PGRNN, yielding $\widebar{\mathcal{M}}_{PGRNN}$:
\begin{align}
\begin{alignedat}{2}
\dot{\hat{\vec{x}}}_{t} &= \,\rec{\widebar{\vec{\mathscr{F}}}}\left( \mathcal{H}^{\hat{\vec{x}}}_{t}, \mathcal{H}^{\vec{u}}_{t}, \mathcal{H}^{t}_{t} ; \vec{\theta}, \tilde{\vec{p}} \right), \, & \rec{\widebar{\vec{\mathscr{F}}}} &: \hat{\mathbb{X}}^{N_t} \times \mathbb{U}^{N_t} \times \mathbb{T}^{N_t} \mapsto \hat{\mathbb{X}},\\
\hat{\vec{y}}_{t} &= \,\rec{\widebar{\vec{\mathscr{G}}}}\left( \mathcal{H}^{\hat{\vec{x}}}_{t}, \mathcal{H}^{\vec{u}}_{t}, \mathcal{H}^{t}_{t} ; \vec{\theta}, \tilde{\vec{p}} \right), \,& \rec{\widebar{\vec{\mathscr{G}}}} &: \hat{\mathbb{X}}^{N_t} \times \mathbb{U}^{N_t} \times \mathbb{T}^{N_t} \mapsto \mathbb{Y}.
\end{alignedat}
\end{align}

\subsection{Physics-Based Constraints}\label{sec:physics_based_constraints}
In control engineering we expect models to be physically plausible. If the dynamics of a given model $\mathcal{M}$ and the corresponding system $\mathcal{S}$ are compared by means of a physics-based restriction, e.g., energy balance at time $t$, and deviate only marginally from each other for all data samples considered, we will call the model $\mathcal{M}$ \emph{physically plausible}. 
Thus, assuming that the physics-based model $\mathcal{M}_{phy}$ with $\tilde{\vec{f}}: \tilde{\mathbb{X}} \times \mathbb{U} \times \mathbb{T} \mapsto \tilde{\mathbb{X}}$ is physically plausible, we only need to ensure that the mapping function $\rec{\vec{\mathscr{F}}}: \hat{\mathbb{X}}^{N_t} \times \mathbb{U}^{N_t} \times \mathbb{T}^{N_t} \times \tilde{\mathbb{X}}^{N_t} \mapsto \hat{\mathbb{X}}$ respects basic physics laws for the whole hybrid model $\widebar{\mathcal{M}}_{PGRNN}$ to be physically plausible as well.
Since data-driven modeling strategies constitute iterative optimization processes, one could suggest to embed the desired physics-based constraints as plain (optimization) constraints on the cost function, e.g., see \cite{Raymond.2021}. However, to conserve the conditioning of the training process, in contrast to previous works, \cite{Dener.2020, Gotte.2022}, a more sophisticated approach is used here.

As introduced in \cite{Kreielmeier.1979}, \emph{vector performance index optimization} 
can be used to simultaneously optimize for several mutually independent objective functions. This multi-objective optimization approach involves the construction of boundaries, one for each objective function, and their successive contraction, yielding Pareto-optimal solutions. Based on their approach, in each training iteration $i$, we define \emph{default values} $c^{(i)}_j\in\left[0,\infty\right)$ with $j = 1,\dots,L$ to norm the $L$ respective objective values $Loss^{(i)}_j > 0$. We obtain the training objective $J^{(i)}$ as the maximum normed objective:
\begin{equation}
\begin{gathered}
J^{(i)}\left(\cdot\right) = \max\left\lbrace \dfrac{Loss^{(i)}_{1}\left(\cdot\right)}{c^{(i)}_{1}}, \ldots, \dfrac{Loss^{(i)}_{L}\left(\cdot\right)}{c^{(i)}_{L}} \right\rbrace,\\
Loss^{(i)}_{j} \leq c^{(i)}_{j} \leq c^{(i-1)}_{j} \leq \ldots \leq c^{(0)}_{j}.
\end{gathered}
\end{equation}	
Through progressive adjustment of the default values, all objective functions are optimized for. Within our work, we have $L=2$ as we consider a typical error-based loss term and a physics-based loss term, latter given in \eqref{eq:energyloss}.

In the following, similar to \cite{Antonelo.2021, Raymond.2021, Gotte.2022}, we use the energy balance as a physics-based restriction. Let $\Delta E_{\mathcal{M}}^{(l)}(t)$ be a model $\mathcal{M}$'s energy discrepancy for sample $l$ at time $t$ with sample length $n^{(l)}$, and let $\Delta E_{\mathcal{M}_{phy}}^{(l)}(t) \overset{!}{=} 0$ be given. Experiments suggest, that an appropriate physics-based objective for a model $\mathcal{M}$ is then
\begin{align}\label{eq:energyloss}
Loss^{(i)}_{energy} := \dfrac{1}{N}\sum_{l=1}^{N}\dfrac{1}{n^{(l)}}\int_{t_0}^{t_{n^{(l)}}} \left| \Delta E_{\mathcal{M}}^{(l, i)}(\tilde{t}) - \Delta E_{\mathcal{S}}^{(l)}(\tilde{t}) \right| \mathrm{d}\tilde{t}.
\end{align}
Here, $N$ is the number of training samples with respective sample lengths $n^{(l)}$, $l=1,\ldots,N$.
Since $\mathcal{S} \neq \mathcal{M}_{phy}$, in order to model the real system $\mathcal{S}$, we enforce $\left|\Delta E_{\mathcal{M},t} - \Delta E_{\mathcal{S},t} \right| \overset{!}{=} 0$, with $\Delta E_{\mathcal{M},t} := \Delta E_{\mathcal{M}}(t)$ and $\Delta E_{\mathcal{S},t} := \Delta E_{\mathcal{S}}(t)$.
The resulting physics-guided model is expected to obey physics-based restrictions through its multi-objective characteristics and is hence referred to as a \emph{Multi-Objective Physics-Guided Recurrent Neural Network}.

\section{Experimental Results}\label{sec:experiments}
To demonstrate the performance of the previously described methods, two nonlinear, non-autonomous systems with increasing model complexity are considered. In a first step, the MOPGRNN's capability of successfully identifying a system's dynamics is investigated simulatively. For this, we are considering a simple, damped pendulum. Subsequently, the setup is generalized to real-data experiments. As an illustrative application, we are modeling the dynamics of a golf-playing robot, which is a test rig at our laboratory and used as a benchmark for various machine learning methods.

For either systems, the physics-guided model $\mathcal{M}_{PGRNN}$ (Sec.~\ref{sec:pgrnn}) and the physics-guided model with constraints $\mathcal{M}_{MOPGRNN}$ (Sec.~\ref{sec:physics_based_constraints}) are compared to a purely data-driven approach, a RNN model $\mathcal{M}_{RNN}$ as described in Sec.~\ref{sec:rnn}, and the respective physics-based model $\mathcal{M}_{phy}$ as a baseline. In this process, both their model accuracy and training data requirements are assessed.

\subsection{Simulative Experiments}\label{sec:simulative_experiments}
\begin{figure}[h]
	\centering
	\subfloat[Visualization]{
		\centering
		\def\svgwidth{.4\columnwidth}	
		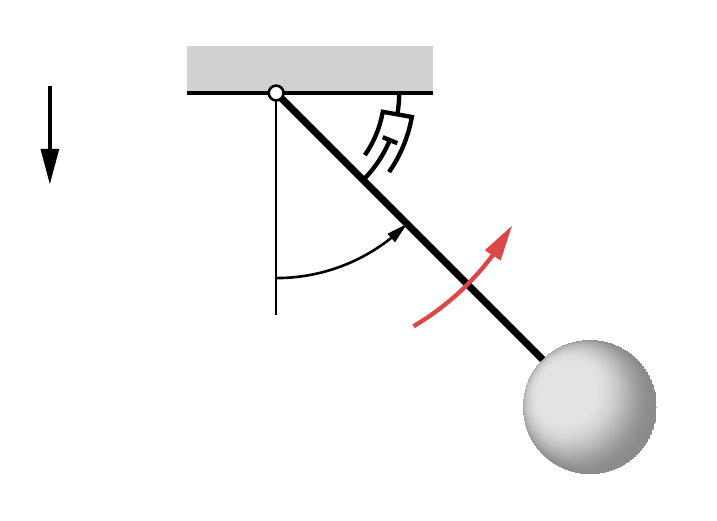
		\label{fig:pendulum}
	}\\
	\subfloat[Parameter sets $\vec{p}$ and $\tilde{\vec{p}}$]{
		\begin{tabular}[b]{lccl}
			
		Parameter & \multicolumn{2}{c}{Value} & \\
			& $\mathcal{S}$ & $\mathcal{M}_{phy}$ & \\
			\hline
			mass $m$ & $50$ & $50$ & $\si{\kilogram}$ \\
			arm length $l$ & $0.045$ & $0.05$ & $\si{\centi\metre}$ \\
			damping constant $d$ & $2.1$ & $2$ & $\si{\kilogram\square\metre\per\second}$ \\
			gravitational constant $g$ & $9.81$ & $9.81$ & $\si{\metre\per\square\second}$\\
			\hline
		\end{tabular}
		\label{tab:parametersPendulum}
	}
	\caption{Damped pendulum}
\end{figure}

The damped pendulum displayed in Fig.~\ref{fig:pendulum} is excited by an external torque $u_t$ and features some friction, which is modeled by a damping constant $d$. Its state $\vec{x}_t=[\varphi_t,\, \dot{\varphi}_t]\in\mathbb{R}^2$ comprises the pendulum's angle $\varphi_t$ as well as the angular velocity $\dot{\varphi}_t$. Both are evolving in time $t$ and are modeled by a system of nonlinear, input affine ODEs. Thus, the ground truth system $\mathcal{S}$ is given by
\begin{equation}\label{eq:pendulum}
\begin{split} 
\dot{\vec{x}}_{t}
= \begin{bmatrix}\dot{\varphi}_t\\(m l^2)^{-1}\left(-m g l\sin\varphi_t - d \dot{\varphi}_t + u_t\right)\end{bmatrix}.
\end{split}
\end{equation}
The physics-based model $\mathcal{M}_{phy}$ is defined analogously for $\tilde{\varphi}_t$ and $\dot{\tilde{\varphi}}_t$. Therefore, the ground truth system $\mathcal{S}$ and the physics-based model $\mathcal{M}_{phy}$ differ solely in terms of their model parameters $\vec{p}$ and $\tilde{\vec{p}}$, which are given in Fig.~\ref{tab:parametersPendulum}.

Data for training is generated by sinusoidal excitation of $\mathcal{S}$ for varying initial states $\vec{x}_0$, a sequence length of $5\si{\second}$ and a sampling rate of $1\si{\kilo\hertz}$. In contrast to real data experiments, the simulation setting enables us to take measurements of the full state. Furthermore, the MOPGRNN's training routine necessitates the derivation of a physics-based constraint \eqref{eq:energyloss}, inciting energy conserving models, which is given by
\begin{equation}
\Delta E_{\mathcal{M},t} = m l^2\ddot{\hat{\varphi}}_t\dot{\hat{\varphi}}_t + m g l\dot{\hat{\varphi}}_t\sin\hat{\varphi}_t + d\dot{\hat{\varphi}}_t^2 - u_t\dot{\hat{\varphi}}_t.
\end{equation}

Before the training process itself, suitable hyperparameters are determined. In particular, the number of hidden neurons is optimized via Bayesian optimization, c.f. \cite{Shahriari.2016}. Once an appropriate number of neurons is found, 16 identical networks are initialized and trained using the ADAM algorithm, as described in \cite{Kingma.2015}, and tested on a distinct test set.

The quality of the obtained models is measured by their simulation error $E_{sim}$, which can be interpreted as the error area between the predicted and target trajectories:
\begin{equation}\label{eq:objective}
E^{(i)}_{sim} = \dfrac{1}{N}\sum_{l=1}^{N}\dfrac{1}{n^{(l)}}\int_{t_0}^{t_{n^{(l)}}}\dfrac{\left|\left| \dot{\vec{x}}^{(l,i)}(\tilde{t}) - \dot{\hat{\vec{x}}}^{(l)}(\tilde{t}) \right|\right|_1}{1 + \lambda \tilde{t}}\mathrm{d}\tilde{t}.
\end{equation}
Here, $N$ is the number of test samples with respective sample lengths $n^{(l)}$, $l=1,\ldots,N$, while $\dot{\hat{\vec{x}}}^{(l,i)}$ and $\dot{\vec{x}}^{(l)}$ are the $l$-th predicted and target trajectories for iteration $i$, respectively. To impose an additional penalty on early deviations, a discount factor $\lambda$ is introduced.
The simulation error obtained for all 16 networks results in an average simulation error and a standard deviation. While the simulation error $E_{sim}$ on the respective data set can be used to determine the validation and testing performance, calculating the gradient of $E_{sim}$ with respect to the network's learnables $\vec{\theta}$ is generally not feasible. Therefore, a mean absolute error loss is used for training.

\begin{wrapfigure}[]{r}{.45\columnwidth}
	\centering
	\def\svgwidth{.45\columnwidth}	
	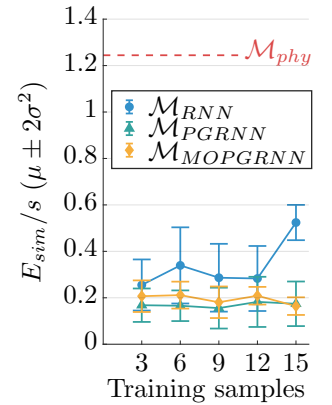
	\caption{Model performance for the damped pendulum}
	\label{fig:results_pendulum}
	\vspace{-10pt}
\end{wrapfigure}
Fig.~\ref{fig:results_pendulum} depicts the average simulation error obtained for the different model types using 3 to 15 training sequences. Since the chosen physics-based model is of low quality due to the corrupted parameters, all three data-based models outperform $\mathcal{M}_{phy}$ as expected by a substantial margin, with no consistent influence of the training data amount on the perceived model accuracy. This could be attributed to both the high sampling rate as well as the simplistic target system.

Both physics-guided hybrid models, e.g., PGRNN and MOPGRNN, demonstrate superior performance over the purely data-driven RNN. In terms of modeling accuracy, the embedded physics-based model causes an additional reduction in simulation error of 34\% to 67\% for the PGRNN and 19\% to 67\% for the MOPGRNN, using the same sequencing of training samples for all experiments. Therefore, apart from a slightly lower standard deviation, the MOPGRNN's accuracy is on par with that of the PGRNN, suggesting that the MOPGRNN's physics-based restriction merely improves training conditioning. This is reflected in more consistent results and faster convergence of the training algorithm.

\subsection{Real-Data Experiments}\label{sec:real_data_experiments}
\begin{figure}[h]
	\centering
	\subfloat[Setup]{\includegraphics[width=0.25\columnwidth]{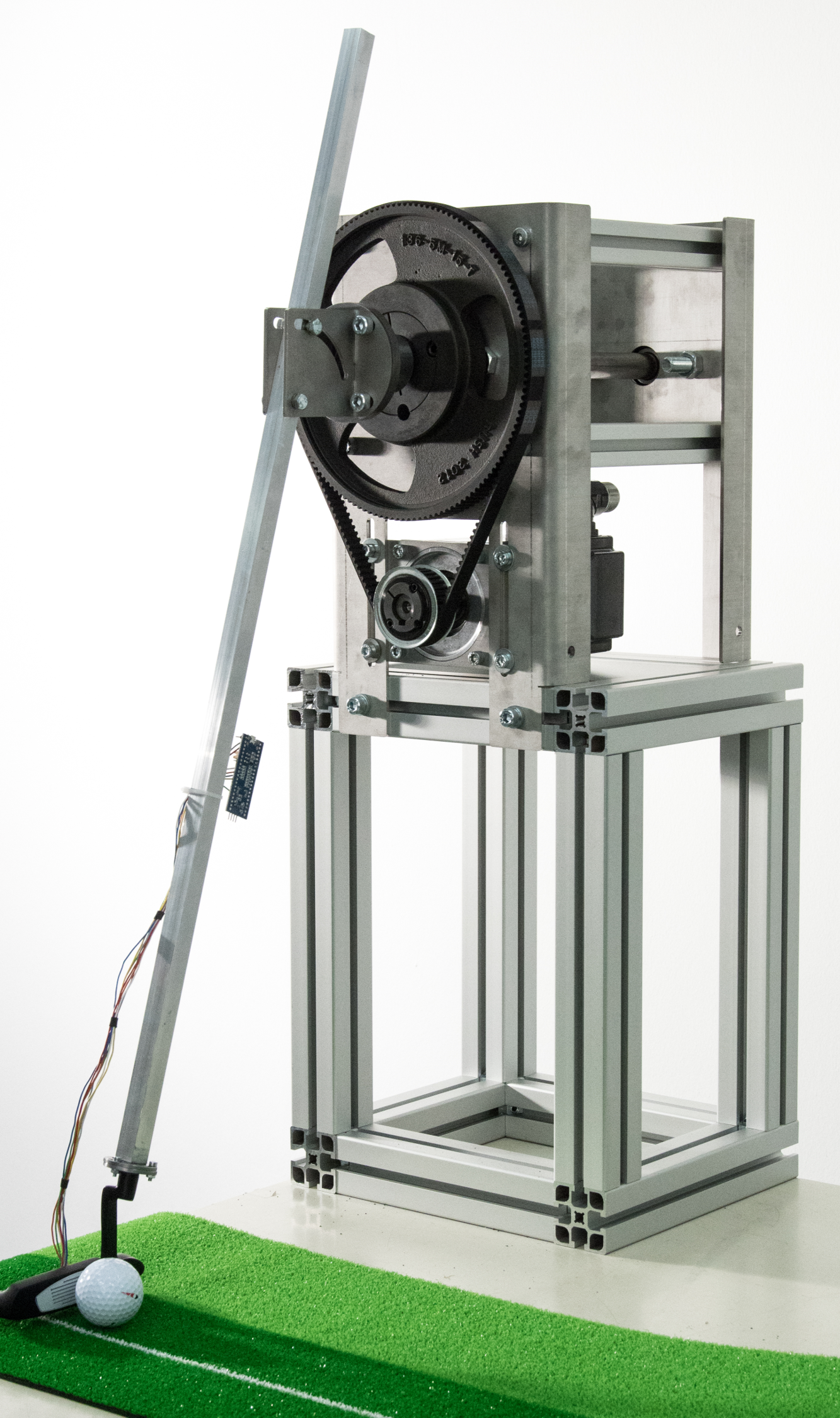}\label{fig:golfrobot}}\\
    \subfloat[Parameters $\tilde{\vec{p}}$]{\begin{tabular}[b]{lrl}
			
			Parameter & Value &\\
			\hline
			mass of club head $m$ & $0.5241$&$\si{\kilogram}$\\
			length from center of &&\\
			gravity to rotation axis $a$ &$0.4702$&$\si{\metre}$\\
			moment of inertia $J$ & $0.1445$&$\si{\kilogram\square\metre}$\\
			damping constant $d$ & $0.0132$&$\si{\kilogram\square\metre\per\second}$\\
			distance of friction point &&\\
			to rotation axis $r$ & $0.0245$&$\si{\metre}$\\
			friction coefficient $\mu$ & $1.5136$&$-$\\
			gravitational constant $g$ & $9.81$&$\si{\metre\per\square\second}$\\
			\hline
		\end{tabular}\label{tab:golfParameters}}
	\caption{Golf robot}
\end{figure}

After yielding promising simulation results, the MOPGRNN is tested on real data from a more complex system. As illustrated in Fig.~\ref{fig:golfrobot}, the golf robot is a test bench at our laboratory, used to study the potential of various machine learning methods on a real-world control problem, see \cite{Junker.2021,Gotte.2022}.

Following extensive modeling efforts, a system of nonlinear, input affine ODEs describing the robot's dynamics has been developed:
\begin{equation}
\begin{gathered}\label{eq:golfrobot}
\dot{\tilde{\vec{x}}}_t
= \begin{bmatrix}\dot{\tilde{\varphi}}_t\\
J^{-1}\left(-m g a\sin\tilde{\varphi}_t-d\dot{\tilde{\varphi}}_t - M_{F,t}+4u_t\right)
\end{bmatrix},\\
M_{F,t} = r\mu\tanh\gamma\dot{\tilde{\varphi}}_t \cdot \left| m \dot{\tilde{\varphi}}_t^{2}a + m g\cos\left(\tilde{\varphi}_t\right)\right|.\\
\end{gathered}
\end{equation}

The motor torque $u_t\in\mathbb{R}$, multiplied by a gear ratio of four, serves as the control input, while the angle $\tilde{\varphi}_t$ and the angular velocity $\dot{\tilde{\varphi}}_t$ of the club head form the robot's model state $\tilde{\vec{x}}_t=[\tilde{\varphi}_t,\, \dot{\tilde{\varphi}}_t]\in\mathbb{R}^2$. The friction term $M_{F,t}$ models the partial dynamics resulting from static and sliding friction, while the parameters $\tilde{\vec{p}} = [m,a,J,d,r,\mu]^T\in\mathbb{R}^6$ shown in Fig.~\ref{tab:golfParameters} have been estimated using particle swarm optimization, c.f. \cite{Kennedy.1995}.

Training is carried out by the same procedure outlined in the previous section. In addition to the performed steps mentioned, the golf robot's measurements need further treatment. Since the chosen approach requires full information of state and state derivative for training and only the club's angle $\varphi_t$ can be measured directly, the angular velocity $\dot{\varphi}_t$ as well as the angular acceleration $\ddot{\varphi}_t$ need to be approximated. In a classical approach, these non-measurable state variables are estimated using a model-based observer, provided a sufficiently accurate system model is available. Since adequate model accuracy cannot be guaranteed here, numerical differentiation is used instead. As data for training, validation, and testing, measurement data for sinusoidal, step, and chirp excitations is captured. Again, a physics-based restriction guaranteeing physical plausibility \eqref{eq:energyloss} is used with the MOPGRNN. Note that the enforced energy balance for the golf robot is based on the parameters of the physics-based model and is given by 
\begin{align}
\begin{split}
\Delta E_{\mathcal{M},t} = J\ddot{\hat{\varphi}}_t\dot{\hat{\varphi}}_t + m g a\dot{\hat{\varphi}}_t\sin\hat{\varphi}_t + d\dot{\hat{\varphi}}_t^2 + M_{F,t} \dot{\hat{\varphi}}_t  - 4\dot{\hat{\varphi}}_t u_{t}.
\end{split}
\end{align}

\begin{figure}[h]
	\centering
	\subfloat[][]{
		\centering
		\def\svgwidth{.45\columnwidth}	
		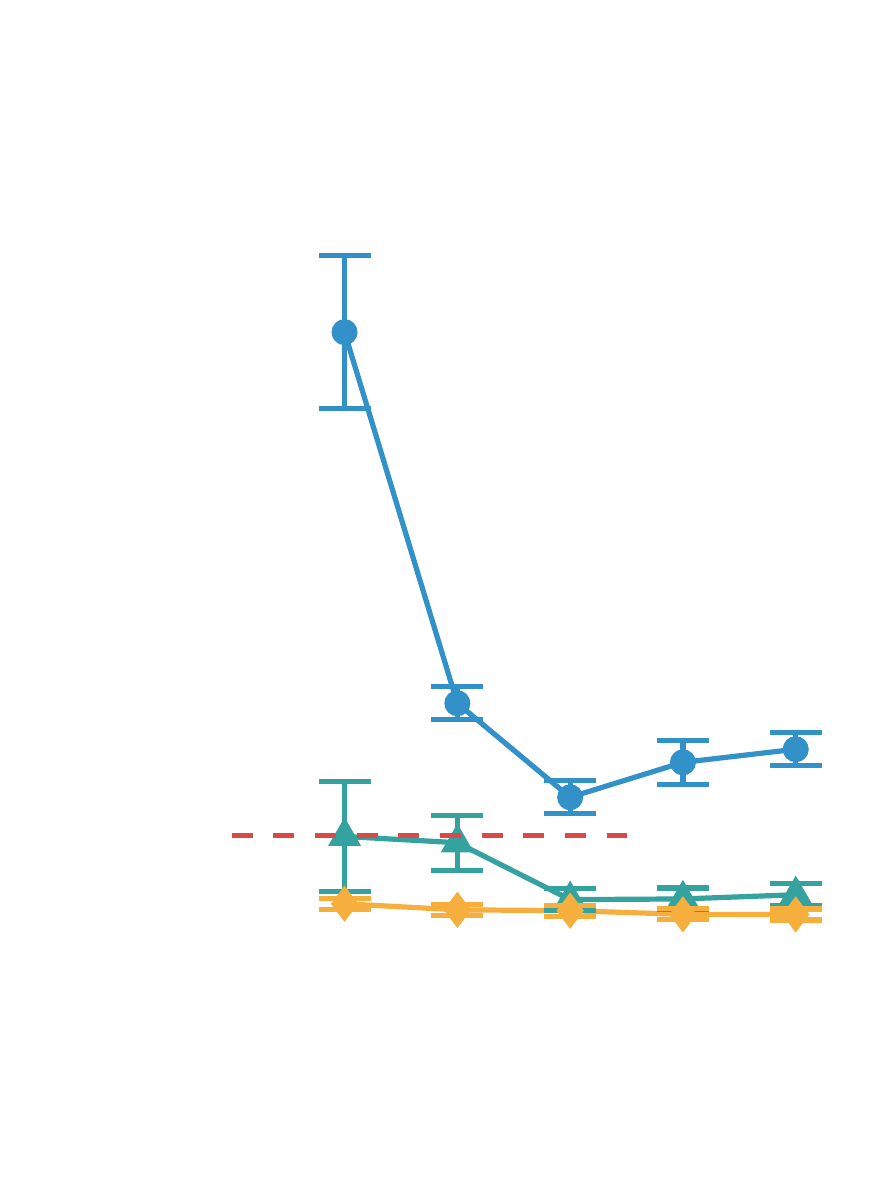
	}
	\hfill
	\subfloat[][]{
		\centering
		\def\svgwidth{.45\columnwidth}	
		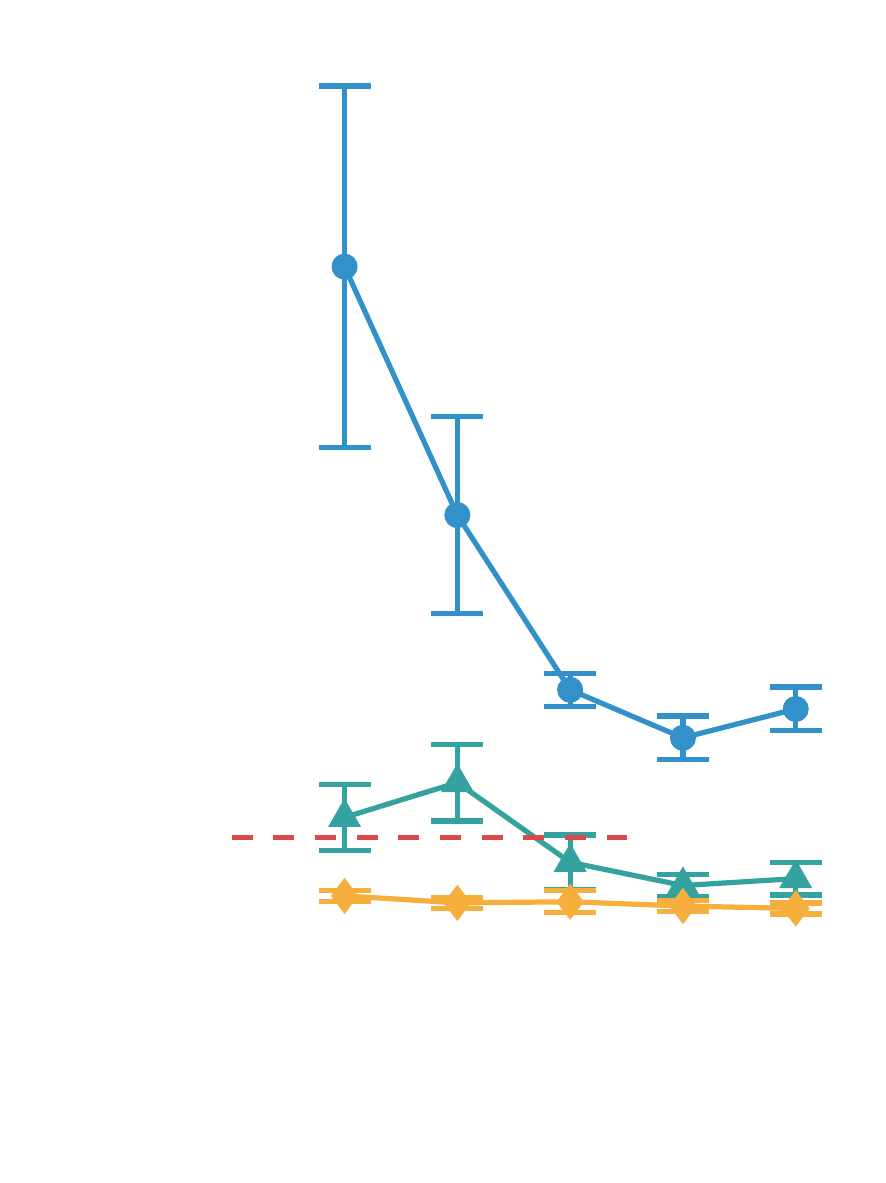
	}
	\caption{Model performance for golf robot (a) without and (b) with sinusoidal training data}\label{fig:golfresults}
\end{figure}

Fig.~\ref{fig:golfresults} shows a comparison of the accuracy of the different model types for varying training data amounts. Due to strong nonlinear effects observed with sinusoidal excitation, both results (a) excluding and (b) including samples with sinusoidal excitation are considered. Analogue to the simulation results obtained in Sec.~\ref{sec:simulative_experiments}, both physics-guided model variants granted higher model accuracy compared to a purely data-driven approach. However, unlike with previous experiments, the RNN's simulation error remained well above the reference error of the physics-based model $\mathcal{M}_{phy}$ with a performance gap of at least 23\%, featuring a steep increase in simulation error for small amounts of training data. Including sinusoidal excitations, the remaining performance gap is even greater, reaching only +59\% compared to the physics-based model.

In contrast to the solely data-driven approach, more encouraging results were observed for the physics-guided models. Starting from approximately one level with the physics-based model, the PGRNN's simulation error drops by up to 38\% ((b) 29\%) for an increasing amount of training data, with a considerably lower standard deviation. By introducing a physics-based restriction with the MOPGRNN, we obtain the best surrogate model of the golf robot observed so far, outperforming the elaborate physics-based model by an additional 46\% and 42\%, respectively. Even for small amounts of training data, hardly any performance degradation is detected. Therefore, a higher level of physics-based knowledge infused into the model correlates with more accurate and consistent results.

We want to emphasize that comparable results can be obtained using a data-driven method like e.g., SINDy by \cite{Brunton.2016}, given an adequate library of candidate functions is identified. However, in contrast to our method, approaches like SINDy do not account for the physical plausibility of the obtained model.

\begin{figure}[h]
	\centering
	\def\svgwidth{\columnwidth}	
	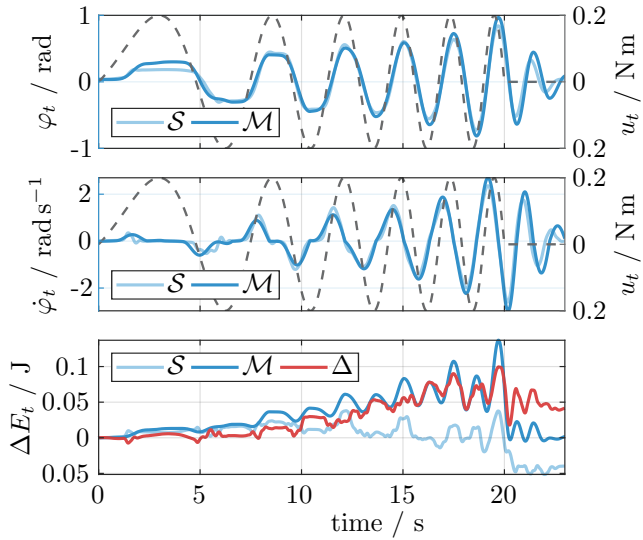
	\caption{Trajectories of simulated $\mathcal{M}_{MOPGRNN}$ and real golf robot $\mathcal{S}$ excited by $u_t$ (in gray dashed) as well as evolution of $\Delta E_{\mathcal{S}, t}$ (in light blue), $\Delta E_{MOPGRNN, t}$ (in blue) and $\Delta$ (in red).}\label{fig:golftrajectories}
\end{figure}

To illustrate the MOPGRNN's simulative performance in terms of accuracy and physical plausibility, both the predicted trajectories and the violation $\Delta := \left|\Delta E_{\mathcal{M},t} - \Delta E_{\mathcal{S},t} \right|$ of the physics-based restriction are shown in Fig.~\ref{fig:golftrajectories}. Except of small deviations, the $\mathcal{M}_{MOPGRNN}$ succeeds at capturing the golf robot's dynamics (in light blue). Despite progressive accumulation, absolute constraint violation (in red) remains reasonably low. Since in practice the obtained model is used in conjunction with a model-based controller with regular feedback, long prediction horizons are usually not relevant anyway.

\section{Conclusion and outlook}\label{sec:conclusion} 
In this paper, a novel framework for the identification of dynamical systems under control is proposed, extending existing control-related techniques by a RNN and a multi-objective strategy yielding physically-plausible hybrid models. The proposed physics-guided model is tested and compared to a conventional physics-based model on both simulative and real data. While a purely data-driven approach failed at identifying a sufficiently accurate model, we were able to outperform the physics-based model by up to 46\%. In combination with an extremely high level of reproducibility, reduced training time, and the guarantee of physical plausibility, the MOPGRNN demonstrates superior performance throughout. Future work include the investigation of different modeling depths on the physics-based prior model, the study of partial observability and the integration with a model-based controller.

\bibstyle{ifacconf}

\end{document}

%% file: Fig1.pdf_tex
\begingroup%
  \makeatletter%
  \providecommand\color[2][]{%
    \errmessage{(Inkscape) Color is used for the text in Inkscape, but the package 'color.sty' is not loaded}%
    \renewcommand\color[2][]{}%
  }%
  \providecommand\transparent[1]{%
    \errmessage{(Inkscape) Transparency is used (non-zero) for the text in Inkscape, but the package 'transparent.sty' is not loaded}%
    \renewcommand\transparent[1]{}%
  }%
  \providecommand\rotatebox[2]{#2}%
  \newcommand*\fsize{\dimexpr\f@size pt\relax}%
  \newcommand*\lineheight[1]{\fontsize{\fsize}{#1\fsize}\selectfont}%
  \ifx\svgwidth\undefined%
    \setlength{\unitlength}{333.06999207bp}%
    \ifx\svgscale\undefined%
      \relax%
    \else%
      \setlength{\unitlength}{\unitlength * \real{\svgscale}}%
    \fi%
  \else%
    \setlength{\unitlength}{\svgwidth}%
  \fi%
  \global\let\svgwidth\undefined%
  \global\let\svgscale\undefined%
  \makeatother%
  \begin{picture}(1,0.38280242)%
    \lineheight{1}%
    \setlength\tabcolsep{0pt}%
    \put(0,0){\includegraphics[width=\unitlength,page=1]{Fig1.pdf}}%
    \put(0.10468145,0.19273531){\color[rgb]{0,0,0}\makebox(0,0)[rt]{\lineheight{1.25}\smash{\begin{tabular}[t]{r}$\mathcal{H}^{\vec{u}}_{t}$\end{tabular}}}}%
    \put(0.10416534,0.23528884){\color[rgb]{0,0,0}\makebox(0,0)[rt]{\lineheight{1.25}\smash{\begin{tabular}[t]{r}$\mathcal{H}^{\hat{\vec{x}}}_{t}$\end{tabular}}}}%
    \put(0.74762835,0.19097659){\color[rgb]{0,0,0}\makebox(0,0)[t]{\lineheight{1.25}\smash{\begin{tabular}[t]{c}$\vec{h}_t$\end{tabular}}}}%
    \put(0.93248385,0.16182228){\color[rgb]{0,0,0}\makebox(0,0)[lt]{\lineheight{1.25}\smash{\begin{tabular}[t]{l}$\dot{\hat{\vec{x}}}_t$\end{tabular}}}}%
    \put(0.67144142,0.16930709){\color[rgb]{0,0,0}\makebox(0,0)[t]{\lineheight{1.25}\smash{\begin{tabular}[t]{c}GRU\end{tabular}}}}%
    \put(0.35843396,0.09498848){\color[rgb]{0,0,0}\makebox(0,0)[t]{\lineheight{1.25}\smash{\begin{tabular}[t]{c}$\mathcal{M}_{phy}$\end{tabular}}}}%
    \put(0.51626379,0.12142905){\color[rgb]{0,0,0}\makebox(0,0)[t]{\lineheight{1.25}\smash{\begin{tabular}[t]{c}$\mathcal{H}^{\dot{\tilde{\vec{x}}}}_{t}$\end{tabular}}}}%
    \put(0,0){\includegraphics[width=\unitlength,page=2]{Fig1.pdf}}%
    \put(0.17603406,0.29855907){\color[rgb]{0,0,0}\makebox(0,0)[t]{\lineheight{1.25}\smash{\begin{tabular}[t]{c}\color{hnilightgreen}(b)\end{tabular}}}}%
    \put(0.52201015,0.29855907){\color[rgb]{0,0,0}\makebox(0,0)[t]{\lineheight{1.25}\smash{\begin{tabular}[t]{c}\color{hnigray}(a)\end{tabular}}}}%
    \put(0.6713421,0.34363517){\color[rgb]{0,0,0}\makebox(0,0)[rt]{\lineheight{1.25}\smash{\begin{tabular}[t]{r}$\mathcal{H}^{t}_{t}$\end{tabular}}}}%
  \end{picture}%
\endgroup%

%% file: Fig2a.pdf_tex
\begingroup%
  \makeatletter%
  \providecommand\color[2][]{%
    \errmessage{(Inkscape) Color is used for the text in Inkscape, but the package 'color.sty' is not loaded}%
    \renewcommand\color[2][]{}%
  }%
  \providecommand\transparent[1]{%
    \errmessage{(Inkscape) Transparency is used (non-zero) for the text in Inkscape, but the package 'transparent.sty' is not loaded}%
    \renewcommand\transparent[1]{}%
  }%
  \providecommand\rotatebox[2]{#2}%
  \newcommand*\fsize{\dimexpr\f@size pt\relax}%
  \newcommand*\lineheight[1]{\fontsize{\fsize}{#1\fsize}\selectfont}%
  \ifx\svgwidth\undefined%
    \setlength{\unitlength}{206.25bp}%
    \ifx\svgscale\undefined%
      \relax%
    \else%
      \setlength{\unitlength}{\unitlength * \real{\svgscale}}%
    \fi%
  \else%
    \setlength{\unitlength}{\svgwidth}%
  \fi%
  \global\let\svgwidth\undefined%
  \global\let\svgscale\undefined%
  \makeatother%
  \begin{picture}(1,0.72727273)%
    \lineheight{1}%
    \setlength\tabcolsep{0pt}%
    \put(0,0){\includegraphics[width=\unitlength,page=1]{Fig2a.pdf}}%
    \put(0.66562077,0.13811553){\color[rgb]{0,0,0}\makebox(0,0)[t]{\lineheight{1.25}\smash{\begin{tabular}[t]{c}$m$\end{tabular}}}}%
    \put(0.61618414,0.41765154){\color[rgb]{0,0,0}\makebox(0,0)[t]{\lineheight{1.25}\smash{\begin{tabular}[t]{c}$l$\end{tabular}}}}%
    \put(0.12340794,0.54436086){\color[rgb]{0,0,0}\makebox(0,0)[t]{\lineheight{1.25}\smash{\begin{tabular}[t]{c}$g$\end{tabular}}}}%
    \put(0.50610816,0.30626453){\color[rgb]{0,0,0}\makebox(0,0)[t]{\lineheight{1.25}\smash{\begin{tabular}[t]{c}$\varphi$\end{tabular}}}}%
    \put(0.71359403,0.43181462){\color[rgb]{0.85882353,0.28235294,0.28235294}\makebox(0,0)[t]{\lineheight{1.25}\smash{\begin{tabular}[t]{c}$u$\end{tabular}}}}%
    \put(0.58208828,0.50770614){\color[rgb]{0,0,0}\makebox(0,0)[lt]{\lineheight{1.25}\smash{\begin{tabular}[t]{l}$d$\end{tabular}}}}%
  \end{picture}%
\endgroup%

%% file: Fig3.pdf_tex
\begingroup%
  \makeatletter%
  \providecommand\color[2][]{%
    \errmessage{(Inkscape) Color is used for the text in Inkscape, but the package 'color.sty' is not loaded}%
    \renewcommand\color[2][]{}%
  }%
  \providecommand\transparent[1]{%
    \errmessage{(Inkscape) Transparency is used (non-zero) for the text in Inkscape, but the package 'transparent.sty' is not loaded}%
    \renewcommand\transparent[1]{}%
  }%
  \providecommand\rotatebox[2]{#2}%
  \newcommand*\fsize{\dimexpr\f@size pt\relax}%
  \newcommand*\lineheight[1]{\fontsize{\fsize}{#1\fsize}\selectfont}%
  \ifx\svgwidth\undefined%
    \setlength{\unitlength}{255bp}%
    \ifx\svgscale\undefined%
      \relax%
    \else%
      \setlength{\unitlength}{\unitlength * \real{\svgscale}}%
    \fi%
  \else%
    \setlength{\unitlength}{\svgwidth}%
  \fi%
  \global\let\svgwidth\undefined%
  \global\let\svgscale\undefined%
  \makeatother%
  \begin{picture}(1,1.33529412)%
    \lineheight{1}%
    \setlength\tabcolsep{0pt}%
    \put(0,0){\includegraphics[width=\unitlength,page=1]{Fig3.pdf}}%
    \put(0.72886029,1.15153412){\makebox(0,0)[lt]{\lineheight{1.25}\smash{\begin{tabular}[t]{l}\textcolor{hnired}{$\mathcal{M}_{phy}$}\end{tabular}}}}%
    \put(0,0){\includegraphics[width=\unitlength,page=2]{Fig3.pdf}}%
    \put(0.38915441,0.10941177){\makebox(0,0)[t]{\lineheight{1.25}\smash{\begin{tabular}[t]{c}3\end{tabular}}}}%
    \put(0.51654412,0.10941177){\makebox(0,0)[t]{\lineheight{1.25}\smash{\begin{tabular}[t]{c}6\end{tabular}}}}%
    \put(0.64393382,0.10941177){\makebox(0,0)[t]{\lineheight{1.25}\smash{\begin{tabular}[t]{c}9\end{tabular}}}}%
    \put(0.77132353,0.10941177){\makebox(0,0)[t]{\lineheight{1.25}\smash{\begin{tabular}[t]{c}12\end{tabular}}}}%
    \put(0.89871324,0.10941177){\makebox(0,0)[t]{\lineheight{1.25}\smash{\begin{tabular}[t]{c}15\end{tabular}}}}%
    \put(0.60147088,0.02313735){\makebox(0,0)[t]{\lineheight{1.25}\smash{\begin{tabular}[t]{c}Training samples\end{tabular}}}}%
    \put(0,0){\includegraphics[width=\unitlength,page=3]{Fig3.pdf}}%
    \put(0.24294118,0.18352941){\makebox(0,0)[rt]{\lineheight{1.25}\smash{\begin{tabular}[t]{r}0\end{tabular}}}}%
    \put(0.24294118,0.33815118){\makebox(0,0)[rt]{\lineheight{1.25}\smash{\begin{tabular}[t]{r}0.2\end{tabular}}}}%
    \put(0.24294118,0.49277324){\makebox(0,0)[rt]{\lineheight{1.25}\smash{\begin{tabular}[t]{r}0.4\end{tabular}}}}%
    \put(0.24294118,0.647395){\makebox(0,0)[rt]{\lineheight{1.25}\smash{\begin{tabular}[t]{r}0.6\end{tabular}}}}%
    \put(0.24294118,0.80201676){\makebox(0,0)[rt]{\lineheight{1.25}\smash{\begin{tabular}[t]{r}0.8\end{tabular}}}}%
    \put(0.24294118,0.95663853){\makebox(0,0)[rt]{\lineheight{1.25}\smash{\begin{tabular}[t]{r}1\end{tabular}}}}%
    \put(0.24294118,1.11126059){\makebox(0,0)[rt]{\lineheight{1.25}\smash{\begin{tabular}[t]{r}1.2\end{tabular}}}}%
    \put(0.24294118,1.26588235){\makebox(0,0)[rt]{\lineheight{1.25}\smash{\begin{tabular}[t]{r}1.4\end{tabular}}}}%
    \put(0.05411765,0.74117706){\rotatebox{90}{\makebox(0,0)[t]{\lineheight{1.25}\smash{\begin{tabular}[t]{c}$E_{sim} / s$ ($\mu\pm2\sigma^2$)\end{tabular}}}}}%
    \put(0,0){\includegraphics[width=\unitlength,page=4]{Fig3.pdf}}%
    \put(0.40994571,0.96434697){\makebox(0,0)[lt]{\lineheight{1.25}\smash{\begin{tabular}[t]{l}$\mathcal{M}_{RNN}$\end{tabular}}}}%
    \put(0,0){\includegraphics[width=\unitlength,page=5]{Fig3.pdf}}%
    \put(0.40994571,0.89228815){\makebox(0,0)[lt]{\lineheight{1.25}\smash{\begin{tabular}[t]{l}$\mathcal{M}_{PGRNN}$\end{tabular}}}}%
    \put(0,0){\includegraphics[width=\unitlength,page=6]{Fig3.pdf}}%
    \put(0.40994571,0.81434697){\makebox(0,0)[lt]{\lineheight{1.25}\smash{\begin{tabular}[t]{l}$\mathcal{M}_{MOPGRNN}$\end{tabular}}}}%
    \put(0,0){\includegraphics[width=\unitlength,page=7]{Fig3.pdf}}%
  \end{picture}%
\endgroup%

%% file: Fig5a.pdf_tex
\begingroup%
  \makeatletter%
  \providecommand\color[2][]{%
    \errmessage{(Inkscape) Color is used for the text in Inkscape, but the package 'color.sty' is not loaded}%
    \renewcommand\color[2][]{}%
  }%
  \providecommand\transparent[1]{%
    \errmessage{(Inkscape) Transparency is used (non-zero) for the text in Inkscape, but the package 'transparent.sty' is not loaded}%
    \renewcommand\transparent[1]{}%
  }%
  \providecommand\rotatebox[2]{#2}%
  \newcommand*\fsize{\dimexpr\f@size pt\relax}%
  \newcommand*\lineheight[1]{\fontsize{\fsize}{#1\fsize}\selectfont}%
  \ifx\svgwidth\undefined%
    \setlength{\unitlength}{255bp}%
    \ifx\svgscale\undefined%
      \relax%
    \else%
      \setlength{\unitlength}{\unitlength * \real{\svgscale}}%
    \fi%
  \else%
    \setlength{\unitlength}{\svgwidth}%
  \fi%
  \global\let\svgwidth\undefined%
  \global\let\svgscale\undefined%
  \makeatother%
  \begin{picture}(1,1.33529412)%
    \lineheight{1}%
    \setlength\tabcolsep{0pt}%
    \put(0,0){\includegraphics[width=\unitlength,page=1]{Fig5a.pdf}}%
    \put(0.72886029,0.38078){\makebox(0,0)[lt]{\lineheight{1.25}\smash{\begin{tabular}[t]{l}\textcolor{hnired}{$\mathcal{M}_{phy}$}\end{tabular}}}}%
    \put(0,0){\includegraphics[width=\unitlength,page=2]{Fig5a.pdf}}%
    \put(0.38915441,0.10941177){\makebox(0,0)[t]{\lineheight{1.25}\smash{\begin{tabular}[t]{c}3\end{tabular}}}}%
    \put(0.51654412,0.10941177){\makebox(0,0)[t]{\lineheight{1.25}\smash{\begin{tabular}[t]{c}6\end{tabular}}}}%
    \put(0.64393382,0.10941177){\makebox(0,0)[t]{\lineheight{1.25}\smash{\begin{tabular}[t]{c}9\end{tabular}}}}%
    \put(0.77132353,0.10941177){\makebox(0,0)[t]{\lineheight{1.25}\smash{\begin{tabular}[t]{c}12\end{tabular}}}}%
    \put(0.89871324,0.10941177){\makebox(0,0)[t]{\lineheight{1.25}\smash{\begin{tabular}[t]{c}15\end{tabular}}}}%
    \put(0.60147088,0.02313735){\makebox(0,0)[t]{\lineheight{1.25}\smash{\begin{tabular}[t]{c}Training samples\end{tabular}}}}%
    \put(0,0){\includegraphics[width=\unitlength,page=3]{Fig5a.pdf}}%
    \put(0.24294118,0.18352941){\makebox(0,0)[rt]{\lineheight{1.25}\smash{\begin{tabular}[t]{r}0\end{tabular}}}}%
    \put(0.24294118,0.33815118){\makebox(0,0)[rt]{\lineheight{1.25}\smash{\begin{tabular}[t]{r}0.05\end{tabular}}}}%
    \put(0.24294118,0.49277324){\makebox(0,0)[rt]{\lineheight{1.25}\smash{\begin{tabular}[t]{r}0.1\end{tabular}}}}%
    \put(0.24294118,0.647395){\makebox(0,0)[rt]{\lineheight{1.25}\smash{\begin{tabular}[t]{r}0.15\end{tabular}}}}%
    \put(0.24294118,0.80201676){\makebox(0,0)[rt]{\lineheight{1.25}\smash{\begin{tabular}[t]{r}0.2\end{tabular}}}}%
    \put(0.24294118,0.95663853){\makebox(0,0)[rt]{\lineheight{1.25}\smash{\begin{tabular}[t]{r}0.25\end{tabular}}}}%
    \put(0.24294118,1.11126059){\makebox(0,0)[rt]{\lineheight{1.25}\smash{\begin{tabular}[t]{r}0.3\end{tabular}}}}%
    \put(0.24294118,1.26588235){\makebox(0,0)[rt]{\lineheight{1.25}\smash{\begin{tabular}[t]{r}0.35\end{tabular}}}}%
    \put(0.05411765,0.74117706){\rotatebox{90}{\makebox(0,0)[t]{\lineheight{1.25}\smash{\begin{tabular}[t]{c}$E_{sim} / s$ ($\mu\pm2\sigma^2$)\end{tabular}}}}}%
    \put(0,0){\includegraphics[width=\unitlength,page=4]{Fig5a.pdf}}%
    \put(0.40854432,1.24212138){\makebox(0,0)[lt]{\lineheight{1.25}\smash{\begin{tabular}[t]{l}$\mathcal{M}_{RNN}$\end{tabular}}}}%
    \put(0,0){\includegraphics[width=\unitlength,page=5]{Fig5a.pdf}}%
    \put(0.40854432,1.17006255){\makebox(0,0)[lt]{\lineheight{1.25}\smash{\begin{tabular}[t]{l}$\mathcal{M}_{PGRNN}$\end{tabular}}}}%
    \put(0,0){\includegraphics[width=\unitlength,page=6]{Fig5a.pdf}}%
    \put(0.40854432,1.09212138){\makebox(0,0)[lt]{\lineheight{1.25}\smash{\begin{tabular}[t]{l}$\mathcal{M}_{MOPGRNN}$\end{tabular}}}}%
    \put(0,0){\includegraphics[width=\unitlength,page=7]{Fig5a.pdf}}%
  \end{picture}%
\endgroup%

%% file: Fig5b.pdf_tex
\begingroup%
  \makeatletter%
  \providecommand\color[2][]{%
    \errmessage{(Inkscape) Color is used for the text in Inkscape, but the package 'color.sty' is not loaded}%
    \renewcommand\color[2][]{}%
  }%
  \providecommand\transparent[1]{%
    \errmessage{(Inkscape) Transparency is used (non-zero) for the text in Inkscape, but the package 'transparent.sty' is not loaded}%
    \renewcommand\transparent[1]{}%
  }%
  \providecommand\rotatebox[2]{#2}%
  \newcommand*\fsize{\dimexpr\f@size pt\relax}%
  \newcommand*\lineheight[1]{\fontsize{\fsize}{#1\fsize}\selectfont}%
  \ifx\svgwidth\undefined%
    \setlength{\unitlength}{255bp}%
    \ifx\svgscale\undefined%
      \relax%
    \else%
      \setlength{\unitlength}{\unitlength * \real{\svgscale}}%
    \fi%
  \else%
    \setlength{\unitlength}{\svgwidth}%
  \fi%
  \global\let\svgwidth\undefined%
  \global\let\svgscale\undefined%
  \makeatother%
  \begin{picture}(1,1.33529412)%
    \lineheight{1}%
    \setlength\tabcolsep{0pt}%
    \put(0,0){\includegraphics[width=\unitlength,page=1]{Fig5b.pdf}}%
    \put(0.72886029,0.37906824){\makebox(0,0)[lt]{\lineheight{1.25}\smash{\begin{tabular}[t]{l}\textcolor{hnired}{$\mathcal{M}_{phy}$}\end{tabular}}}}%
    \put(0,0){\includegraphics[width=\unitlength,page=2]{Fig5b.pdf}}%
    \put(0.38915441,0.10941177){\makebox(0,0)[t]{\lineheight{1.25}\smash{\begin{tabular}[t]{c}3\end{tabular}}}}%
    \put(0.51654412,0.10941177){\makebox(0,0)[t]{\lineheight{1.25}\smash{\begin{tabular}[t]{c}6\end{tabular}}}}%
    \put(0.64393382,0.10941177){\makebox(0,0)[t]{\lineheight{1.25}\smash{\begin{tabular}[t]{c}9\end{tabular}}}}%
    \put(0.77132353,0.10941177){\makebox(0,0)[t]{\lineheight{1.25}\smash{\begin{tabular}[t]{c}12\end{tabular}}}}%
    \put(0.89871324,0.10941177){\makebox(0,0)[t]{\lineheight{1.25}\smash{\begin{tabular}[t]{c}15\end{tabular}}}}%
    \put(0.60147088,0.02313735){\makebox(0,0)[t]{\lineheight{1.25}\smash{\begin{tabular}[t]{c}Training samples\end{tabular}}}}%
    \put(0,0){\includegraphics[width=\unitlength,page=3]{Fig5b.pdf}}%
    \put(0.24294118,0.18352941){\makebox(0,0)[rt]{\lineheight{1.25}\smash{\begin{tabular}[t]{r}0\end{tabular}}}}%
    \put(0.24294118,0.33815118){\makebox(0,0)[rt]{\lineheight{1.25}\smash{\begin{tabular}[t]{r}0.05\end{tabular}}}}%
    \put(0.24294118,0.49277324){\makebox(0,0)[rt]{\lineheight{1.25}\smash{\begin{tabular}[t]{r}0.1\end{tabular}}}}%
    \put(0.24294118,0.647395){\makebox(0,0)[rt]{\lineheight{1.25}\smash{\begin{tabular}[t]{r}0.15\end{tabular}}}}%
    \put(0.24294118,0.80201676){\makebox(0,0)[rt]{\lineheight{1.25}\smash{\begin{tabular}[t]{r}0.2\end{tabular}}}}%
    \put(0.24294118,0.95663853){\makebox(0,0)[rt]{\lineheight{1.25}\smash{\begin{tabular}[t]{r}0.25\end{tabular}}}}%
    \put(0.24294118,1.11126059){\makebox(0,0)[rt]{\lineheight{1.25}\smash{\begin{tabular}[t]{r}0.3\end{tabular}}}}%
    \put(0.24294118,1.26588235){\makebox(0,0)[rt]{\lineheight{1.25}\smash{\begin{tabular}[t]{r}0.35\end{tabular}}}}%
    \put(0.05411765,0.74117706){\rotatebox{90}{\makebox(0,0)[t]{\lineheight{1.25}\smash{\begin{tabular}[t]{c}$E_{sim} / s$ ($\mu\pm2\sigma^2$)\end{tabular}}}}}%
  \end{picture}%
\endgroup%

%% file: Fig6.pdf_tex
\begingroup%
  \makeatletter%
  \providecommand\color[2][]{%
    \errmessage{(Inkscape) Color is used for the text in Inkscape, but the package 'color.sty' is not loaded}%
    \renewcommand\color[2][]{}%
  }%
  \providecommand\transparent[1]{%
    \errmessage{(Inkscape) Transparency is used (non-zero) for the text in Inkscape, but the package 'transparent.sty' is not loaded}%
    \renewcommand\transparent[1]{}%
  }%
  \providecommand\rotatebox[2]{#2}%
  \newcommand*\fsize{\dimexpr\f@size pt\relax}%
  \newcommand*\lineheight[1]{\fontsize{\fsize}{#1\fsize}\selectfont}%
  \ifx\svgwidth\undefined%
    \setlength{\unitlength}{425.25bp}%
    \ifx\svgscale\undefined%
      \relax%
    \else%
      \setlength{\unitlength}{\unitlength * \real{\svgscale}}%
    \fi%
  \else%
    \setlength{\unitlength}{\svgwidth}%
  \fi%
  \global\let\svgwidth\undefined%
  \global\let\svgscale\undefined%
  \makeatother%
  \begin{picture}(1,0.82892416)%
    \lineheight{1}%
    \setlength\tabcolsep{0pt}%
    \put(0,0){\includegraphics[width=\unitlength,page=1]{Fig6.pdf}}%
    \put(0.11216931,0.58553792){\makebox(0,0)[lt]{\lineheight{1.25}\smash{\begin{tabular}[t]{l}-1\end{tabular}}}}%
    \put(0.12275132,0.68518519){\makebox(0,0)[lt]{\lineheight{1.25}\smash{\begin{tabular}[t]{l}0\end{tabular}}}}%
    \put(0.12275132,0.78483245){\makebox(0,0)[lt]{\lineheight{1.25}\smash{\begin{tabular}[t]{l}1\end{tabular}}}}%
    \put(0.08218695,0.69654488){\rotatebox{90}{\makebox(0,0)[t]{\lineheight{1.25}\smash{\begin{tabular}[t]{c}$\varphi_t$ / $\si{\radian}$\end{tabular}}}}}%
    \put(0,0){\includegraphics[width=\unitlength,page=2]{Fig6.pdf}}%
    \put(0.85784832,0.58553792){\makebox(0,0)[lt]{\lineheight{1.25}\smash{\begin{tabular}[t]{l}0.2\end{tabular}}}}%
    \put(0.85784832,0.68518519){\makebox(0,0)[lt]{\lineheight{1.25}\smash{\begin{tabular}[t]{l}0\end{tabular}}}}%
    \put(0.85784832,0.78483245){\makebox(0,0)[lt]{\lineheight{1.25}\smash{\begin{tabular}[t]{l}0.2\end{tabular}}}}%
    \put(0.94955908,0.69285784){\rotatebox{90}{\makebox(0,0)[t]{\lineheight{1.25}\smash{\begin{tabular}[t]{c}$u_t$ / $\si{\newton\metre}$\end{tabular}}}}}%
    \put(0,0){\includegraphics[width=\unitlength,page=3]{Fig6.pdf}}%
    \put(0.25573192,0.62610229){\makebox(0,0)[lt]{\lineheight{1.25}\smash{\begin{tabular}[t]{l}$\mathcal{S}$\end{tabular}}}}%
    \put(0,0){\includegraphics[width=\unitlength,page=4]{Fig6.pdf}}%
    \put(0.36684303,0.62610229){\makebox(0,0)[lt]{\lineheight{1.25}\smash{\begin{tabular}[t]{l}$\mathcal{M}$\end{tabular}}}}%
    \put(0,0){\includegraphics[width=\unitlength,page=5]{Fig6.pdf}}%
    \put(0.11216931,0.37626914){\makebox(0,0)[lt]{\lineheight{1.25}\smash{\begin{tabular}[t]{l}-2\end{tabular}}}}%
    \put(0.12275132,0.4465619){\makebox(0,0)[lt]{\lineheight{1.25}\smash{\begin{tabular}[t]{l}0\end{tabular}}}}%
    \put(0.12275132,0.51685467){\makebox(0,0)[lt]{\lineheight{1.25}\smash{\begin{tabular}[t]{l}2\end{tabular}}}}%
    \put(0.08218695,0.453194){\rotatebox{90}{\makebox(0,0)[t]{\lineheight{1.25}\smash{\begin{tabular}[t]{c}$\dot{\varphi}_t$ / $\si{\radian\per\second}$\end{tabular}}}}}%
    \put(0,0){\includegraphics[width=\unitlength,page=6]{Fig6.pdf}}%
    \put(0.85784832,0.34215168){\makebox(0,0)[lt]{\lineheight{1.25}\smash{\begin{tabular}[t]{l}0.2\end{tabular}}}}%
    \put(0.85784832,0.44179894){\makebox(0,0)[lt]{\lineheight{1.25}\smash{\begin{tabular}[t]{l}0\end{tabular}}}}%
    \put(0.85784832,0.54144621){\makebox(0,0)[lt]{\lineheight{1.25}\smash{\begin{tabular}[t]{l}0.2\end{tabular}}}}%
    \put(0.94955908,0.45296794){\rotatebox{90}{\makebox(0,0)[t]{\lineheight{1.25}\smash{\begin{tabular}[t]{c}$u_t$ / $\si{\newton\metre}$\end{tabular}}}}}%
    \put(0,0){\includegraphics[width=\unitlength,page=7]{Fig6.pdf}}%
    \put(0.25573192,0.38271605){\makebox(0,0)[lt]{\lineheight{1.25}\smash{\begin{tabular}[t]{l}$\mathcal{S}$\end{tabular}}}}%
    \put(0,0){\includegraphics[width=\unitlength,page=8]{Fig6.pdf}}%
    \put(0.36684303,0.38271605){\makebox(0,0)[lt]{\lineheight{1.25}\smash{\begin{tabular}[t]{l}$\mathcal{M}$\end{tabular}}}}%
    \put(0,0){\includegraphics[width=\unitlength,page=9]{Fig6.pdf}}%
    \put(0.15228641,0.06631393){\makebox(0,0)[t]{\lineheight{1.25}\smash{\begin{tabular}[t]{c}0\end{tabular}}}}%
    \put(0.30303851,0.06631393){\makebox(0,0)[t]{\lineheight{1.25}\smash{\begin{tabular}[t]{c}5\end{tabular}}}}%
    \put(0.45162029,0.06631393){\makebox(0,0)[t]{\lineheight{1.25}\smash{\begin{tabular}[t]{c}10\end{tabular}}}}%
    \put(0.60459076,0.06631393){\makebox(0,0)[t]{\lineheight{1.25}\smash{\begin{tabular}[t]{c}15\end{tabular}}}}%
    \put(0.75552772,0.06631393){\makebox(0,0)[t]{\lineheight{1.25}\smash{\begin{tabular}[t]{c}20\end{tabular}}}}%
    \put(0.49870293,0.02516173){\makebox(0,0)[lt]{\lineheight{1.25}\smash{\begin{tabular}[t]{l}time / $\si{\second}$\end{tabular}}}}%
    \put(0,0){\includegraphics[width=\unitlength,page=10]{Fig6.pdf}}%
    \put(0.1397015,0.098606){\makebox(0,0)[rt]{\lineheight{1.25}\smash{\begin{tabular}[t]{r}0.05\end{tabular}}}}%
    \put(0.13997478,0.15195503){\makebox(0,0)[rt]{\lineheight{1.25}\smash{\begin{tabular}[t]{r}0\end{tabular}}}}%
    \put(0.13978188,0.20530423){\makebox(0,0)[rt]{\lineheight{1.25}\smash{\begin{tabular}[t]{r}0.05\end{tabular}}}}%
    \put(0.13979795,0.25865344){\makebox(0,0)[rt]{\lineheight{1.25}\smash{\begin{tabular}[t]{r}0.1\end{tabular}}}}%
    \put(0.05044092,0.20869986){\rotatebox{90}{\makebox(0,0)[t]{\lineheight{1.25}\smash{\begin{tabular}[t]{c}$\Delta E_t$ / $\si{\joule}$\end{tabular}}}}}%
    \put(0,0){\includegraphics[width=\unitlength,page=11]{Fig6.pdf}}%
    \put(0.25573192,0.25925926){\makebox(0,0)[lt]{\lineheight{1.25}\smash{\begin{tabular}[t]{l}$\mathcal{S}$\end{tabular}}}}%
    \put(0,0){\includegraphics[width=\unitlength,page=12]{Fig6.pdf}}%
    \put(0.36684303,0.25925926){\makebox(0,0)[lt]{\lineheight{1.25}\smash{\begin{tabular}[t]{l}$\mathcal{M}$\end{tabular}}}}%
    \put(0,0){\includegraphics[width=\unitlength,page=13]{Fig6.pdf}}%
    \put(0.49911817,0.25925926){\makebox(0,0)[lt]{\lineheight{1.25}\smash{\begin{tabular}[t]{l}$\Delta$\end{tabular}}}}%
    \put(0,0){\includegraphics[width=\unitlength,page=14]{Fig6.pdf}}%
  \end{picture}%
\endgroup%

%% file: Identifying_Non-autonomous_Dynamical_System_with_Physics-Guided_Recurrent_Neural_Networks.bbl
\begin{thebibliography}{15}
 	\providecommand{\natexlab}[1]{#1}
 	\providecommand{\url}[1]{\texttt{#1}}
 	\providecommand{\urlprefix}{URL }
 	\expandafter\ifx\csname urlstyle\endcsname\relax
 	\providecommand{\doi}[1]{doi:\discretionary{}{}{}#1}\else
 	\providecommand{\doi}{doi:\discretionary{}{}{}\begingroup
 		\urlstyle{rm}\Url}\fi
 	
 	\bibitem[{Antonelo et~al.(2021)Antonelo, Camponogara, Seman, Souza, Jordanou, and
 		Hubner}]{Antonelo.2021}
 	Antonelo, E.A., Camponogara, E., Seman, L.O., de Souza, E.R., Jordanou, J.P.,
 	and Hubner, J.F. (2021).
 	\newblock Physics-informed neural nets-based control.
 	
 	\bibitem[{Bishop(2006)}]{Bishop.2006}
 	Bishop, C.M. (2006).
 	\newblock\emph{Pattern recognition and machine learning}.
 	\newblock Springer, New York, NY.
 	
 	\bibitem[{Brunton et~al.(2016)Brunton, Proctor, and Kutz}]{Brunton.2016}
 	Brunton, S.L., Proctor, J.L., and Kutz, J.N. (2016).
 	\newblock Sparse identification of nonlinear dynamics with control (SINDYc).
 	\newblock In \emph{NOLCOS}, volume 49,
 	issue 18, 710--715.
 	
 	\bibitem[{Cho et~al.(2014)Cho, {van Merrienboer}, Gulcehre, Bahdanau, Bougares,
 		Schwenk, and Bengio}]{Cho.2014}
 	Cho, K., {van Merrienboer}, B., Gulcehre, C., Bahdanau, D., Bougares, F.,
 	Schwenk, H., and Bengio, Y. (2014).
 	\newblock Learning phrase representations using RNN encoder-decoder for
 	statistical machine translation.
 	
 	\bibitem[{Dener et~al.(2020)Dener, Miller, Churchill, Munson, and
 		Chang}]{Dener.2020}
 	Dener, A., Miller, M.A., Churchill, R.M., Munson, T., and Chang, C.S. (2020).
 	\newblock Training neural networks under physical constraints using a
 	stochastic augmented Lagrangian approach.
 	\newblock In \emph{Preprint Journal of Computational Physics}.
 	
 	\bibitem[{Garcez and Lamb(2020)}]{Garcez.2020}
 	Garcez, A. and Lamb, L.C. (2020).
 	\newblock Neurosymbolic AI: The 3rd wave.
 	
 	\bibitem[{G{\"o}tte and Timmermann(2022)}]{Gotte.2022}
 	G{\"o}tte, R.-S. and Timmermann, J. (2022).
 	\newblock Composed physics- and data-driven system identification for
 	non-autonomous systems in control engineering.
 	\newblock In \emph{Accepted for AIRC}.
 	
 	\bibitem[{Jia et~al.(2020)Jia, Willard, Karpatne, Read, Zwart, Steinbach, and
 		Kumar}]{Jia.2020}
 	Jia, X., Willard, J., Karpatne, A., Read, J.S., Zwart, J.A., Steinbach, M., and
 	Kumar, V. (2020).
 	\newblock Physics-guided machine learning for scientific discovery: An
 	application in simulating lake temperature profiles.
 	\newblock In \emph{ACM/IMS Transactions on Data Science}, volume 2, issue 3,
 	1--26.
 	
 	\bibitem[{Junker et~al.(2021)Junker, Timmermann, and Tr{\"{a}}chtler}]{Junker.2021}
 	Junker, A., Timmermann, J., and Tr{\"{a}}chtler, A. (2021).
 	\newblock Data-driven models for control engineering applications using the Koopman operator.
 	\newblock In \emph{Accepted for ALCOS}.
 	
 	\bibitem[{Karpatne et~al.(2017)Karpatne, Atluri, Faghmous, Steinbach, Banerjee,
 		Ganguly, Shekhar, Samatova, and Kumar}]{Karpatne.2017}
 	Karpatne, A., Atluri, G., Faghmous, J., Steinbach, M., Banerjee, A., Ganguly,
 	A., Shekhar, S., Samatova, N., and Kumar, V. (2017).
 	\newblock Theory-guided data science: A new paradigm for scientific discovery
 	from data.
 	\newblock \emph{IEEE Transactions on Knowledge and Data Engineering}, 29(10),
 	2318--2331.
 	
 	\bibitem[{Karpatne et~al.(2017)Karpatne, Watkins, Read, and Kumar}]{Karpatne.2017b}
 	Karpatne, A., Watkins, W., Read, J., and Kumar, V. (2017).
 	\newblock Physics-guided neural networks (PGNN): An application in lake
 	temperature modeling.
 	
 	\bibitem[{Kennedy and Eberhart(1995)}]{Kennedy.1995}
 	Kennedy, J. and Eberhart, R. (1995).
 	\newblock Particle swarm optimization.
 	\newblock In \emph{Proceedings of the IJCNN},
 	volume 4, 1942--1948.
 	
 	\bibitem[{Kingma and Ba(2015)}]{Kingma.2015}
 	Kingma, D.P. and Ba, J. (2015).
 	\newblock Adam: A method for stochastic optimization.
 	\newblock In \emph{ICLR}.
 	
 	\bibitem[{Krei{\ss}elmeier and Steinhauser(1979)}]{Kreielmeier.1979}
 	Krei{\ss}elmeier, G. and Steinhauser, R. (1979).
 	\newblock Systematic control design by optimizing a vector performance
 	index.
 	\newblock \emph{IFAC Proceedings Volumes}, volume 12, 113--117.
 	
 	\bibitem[{Raymond and Camarillo(2021)}]{Raymond.2021}
 	Raymond, S.J. and Camarillo, D.B. (2021).
 	\newblock Applying physics-based loss functions to neural networks for improved
 	generalizability in mechanics problems.
 	\newblock In \emph{Preprint Journal of Computational Physics}.
 	
 	\bibitem[{Shahriari et~al.(2016)Shahriari, Swersky, Wang, Adams, and
 		de~Freitas}]{Shahriari.2016}
 	Shahriari, B., Swersky, K., Wang, Z., Adams, R.P., and de~Freitas, N. (2016).
 	\newblock Taking the human out of the loop: A review of Bayesian optimization.
 	\newblock In \emph{Proceedings of the IEEE}, volume 104, 148--175.
 	
 	\bibitem[{Willard et~al.(2021)Willard, Jia, Xu, Steinbach, and
 		Kumar}]{Willard.2021}
 	Willard, J., Jia, X., Xu, S., Steinbach, M., and Kumar, V. (2021).
 	\newblock Integrating scientific knowledge with machine learning for
 	engineering and environmental systems.
 \end{thebibliography}


\begin{thebibliography}{4}
\providecommand{\natexlab}[1]{#1}
\providecommand{\url}[1]{\texttt{#1}}
\providecommand{\urlprefix}{URL }
\expandafter\ifx\csname urlstyle\endcsname\relax
  \providecommand{\doi}[1]{doi:\discretionary{}{}{}#1}\else
  \providecommand{\doi}{doi:\discretionary{}{}{}\begingroup
  \urlstyle{rm}\Url}\fi

\bibitem[{Able(1956)}]{Abl:56}
Able, B. (1956).
\newblock Nucleic acid content of microscope.
\newblock \emph{Nature}, 135, 7--9.

\bibitem[{Able et~al.(1954)Able, Tagg, and Rush}]{AbTaRu:54}
Able, B., Tagg, R., and Rush, M. (1954).
\newblock Enzyme-catalyzed cellular transanimations.
\newblock In A.~Round (ed.), \emph{Advances in Enzymology}, volume~2, 125--247.
  Academic Press, New York, 3rd edition.

\bibitem[{Keohane(1958)}]{Keo:58}
Keohane, R. (1958).
\newblock \emph{Power and Interdependence: World Politics in Transitions}.
\newblock Little, Brown \& Co., Boston.

\bibitem[{Powers(1985)}]{Pow:85}
Powers, T. (1985).
\newblock Is there a way out?
\newblock \emph{Harpers}, 35--47.

\end{thebibliography}
